\documentclass[10pt, twocolumn, a4paper]{scrartcl}
\usepackage[utf8]{inputenc}
\usepackage[T1]{fontenc}
\usepackage{color}	
\usepackage{colortbl}
\usepackage{graphicx,amsmath}
\usepackage{mathptmx,courier}
\usepackage[scaled]{helvet}
\usepackage[ngerman]{babel}
\usepackage{fancyhdr}
\usepackage[bf]{caption2}
\usepackage{upgreek}
\usepackage{units}
\usepackage[numbers]{natbib}
\usepackage{tocstyle}
\usepackage{multicol}
\usepackage{lipsum}
\usepackage{booktabs}

\newcommand{\tsub}[1]{$T_{\text{S}}={#1}\,^{\circ}$C}
\newcommand{\BEP}[3]{$p_{\text{#1}} = {#2} \times 10^{{#3}}$ Torr}
\newcommand{\C}{$^\circ$C }

\newcommand{\kapitell}[1]{\subsubsection*{\textit{#1}}}

\addto\captionsngerman{}

\addto\captionsngerman{

} 
\addto\captionsngerman{

} 

\usetocstyle{allwithdot} 
\begin {document}
\definecolor{dunkelgrau}{rgb}{0.8,0.8,0.8}
\definecolor{hellgrau}{rgb}{0.95,0.95,0.95}

\twocolumn[
  \begin{@twocolumnfalse}
    \begin{abstract}
\begin{center}
\section*{\center{Focused ion beam induced growth of monocrystalline InAs nanowires}}
\end{center}
\begin{center}
\large{\textit{\textbf{S. Scholz}$^{1,}$\footnotemark, R. Schott$^1$, P. A. Labud$^1$, C. Somsen$^2$, D. Reuter$^{1,}\footnotemark$,  A. Ludwig$^1$ and A. D. Wieck$^1$}}
\end{center}
\begin{center}
\large{\textit{$^1$Ruhr-Universität Bochum, Lehrstuhl für Angewandte Festkörperphysik, D-44780 Bochum}}\\
\large{\textit{$^2$Ruhr-Universität Bochum, Lehrstuhl für Werkstoffwissenschaft, D-44780 Bochum}}\\
\end{center}
We investigate monocrystalline InAs nanowires (NWs) which are grown by molecular beam epitaxy (MBE) and induced by focused ion beam (FIB) implanted Au spots. With this unique combination of methods an increase of the aspect ratio, i.e. the length to width ratio, of the grown NWs up to 300 was achieved. To control the morphology and crystalline structure of the NWs, the growth parameters like temperature, flux ratios and implantation fluence are varied and optimized. Furthermore, the influence of the used molecular arsenic species, in particular the As$_2$ to As$_4$ ratio, is investigated and adjusted. In addition to the high aspect ratio, this optimization results in the growth of monocrystalline InAs NWs with a negligible number of stacking faults. Single NWsx were placed site-controlled by FIB implantation, which supplements the working field of area growth. 
    \end{abstract}
  \end{@twocolumnfalse}
]

\begin{figure}[tb]
\includegraphics[width=0.49 \textwidth]{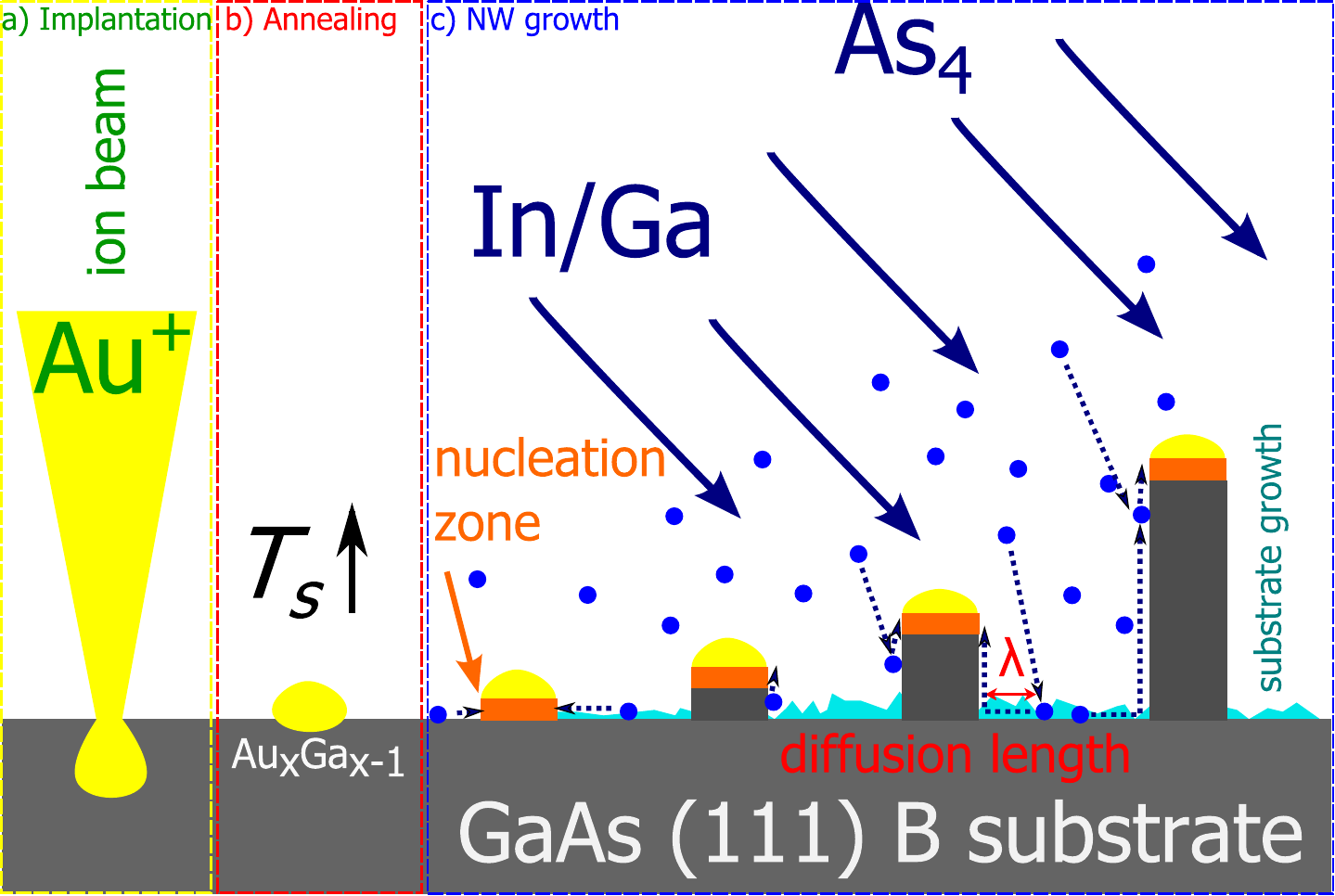}
\captionof{figure}{Schematic of the in-situ sample preparation steps and VSS growth. While the implantation in step a) is done in the FIB the annealing and catalyst creation (b)) and the NW growth (c)) is done in the MBE growth chamber.}
\label{fig:vls}
\end{figure}

\kapitell{I. Introduction}
One-dimensional semiconductor NWs have been attracting growing attention in the last decades, due to their unique physical properties compared to the bulk crystal and superiority as building blocks for nanoscale devices. With application in fields like photovoltaics \cite{Garnett.2010,Kailuweit.2010}, optoelectronics \cite{Yan.2009,Saxena.2013}, single electron transistors \cite{Law.2004,Cui.2001} and nanomechanics \cite{Wu.2005}, NWs were studied by many groups. \\
While it is challenging to use self assembled NWs for further applications, site controlled NWs enable new sample designs which can be of use for electrical contacting the NWs or for setting up photonic crystals. Due to the small diameter, the grown NWs might act as quantum wires which could be examined with regard to the behavior of a one dimensional electron gas (1 DEG).\\ 
\footnotetext[1]{sven.scholz@ruhr-uni-bochum.de}
\footnotetext[2]{Current adress: Universität Paderborn, Arbeitsgruppe für optoelektronische Materialien und Bauelemente, D-33098 Paderborn}\\
There are different methods to induce the NW growth, which can be separated in catalyst and catalyst free mechanisms. In this work, the NW growth was achieved due to the vapor-liquid-solid- (VLS-) or vapor-solid-solid- (VSS-) growth mechanism \cite{Mohammad.2008,Harmand.2005,Givargizov.1975}. These mechanisms are based on a catalyst droplet, which induces the NW growth on the substrate when expose to the molecular beam like shown in Fig.\ref{fig:vls}b)-c).
\\
NWs can be made from different semiconducting materials and nearly arbitrary alloys. In this work, InAs NWs are the main subject of investigation.
InAs NWs are selfconducting and have a high charge carrier mobility, a small band gap and a small effective mass, which is based on the properties of the direct semiconductor InAs. These properties make the InAs nanowires a promising investigation to enhance several applications. \\ Usually, the crystal structure of the MBE grown InAs NWs is difficult to control. Often, a repeatedly change of the crystal structure between zincblende and wurtzite within one NW was reported \cite{Martelli.2011,Zhang.2009}.  This leads to undesired scattering in electrical transport and lower optical qualities of the NWs \cite{Heiss.2011,Chen.2001}. Therefore, the growth of monocrystalline NWs is highly preferable to overcome these disadvantages. \\
The site controlled growth of NWs is mostly done by nanoimprint which necessitates an advanced sample preparation including the fabrication and usage of a mask \cite{Fan.2006,Hersee.2006}. \\ 
A new approach to create the catalyst for the VLS- or VSS-growth mechanism of the NWs  by using FIB implantation is reported in this work.
\begin{table*}[tbh]
\centering

\caption{Implantation and growth parameter for the presented samples. Additionally each sample had a growth time of one hour. }

\begin{tabular}[t]{llllll}
\toprule 
Sample & Implantation fluence & As & In & Ga & Temp. \\ 
\midrule 
Series A & 1 $\cdot$ $10^{14}$ to 2 $\cdot$ $10^{16} \,\,\nicefrac{\text{ions}}{\text{cm}^2}$ & $5.5 \times 10^{-6}$ Torr& $1.8 \times 10^{-7}$ Torr&---& $370^{\circ}$C \\ 
Series B & 1 $\cdot$ $10^{4}$ to 7 $\cdot$ $10^{7} \,\,\nicefrac{\text{ions}}{\text{point}}$ & $5.5 \times 10^{-6}$ Torr& $2.0 \times 10^{-7}$ Torr&---& $370^{\circ}$C \\ 
C & 4 $\cdot$ $10^{15} \,\,\nicefrac{\text{ions}}{\text{cm}^2}$ & $5.5 \times 10^{-6}$ Torr& $1.8 \times 10^{-7}$ Torr&---& $370^{\circ}$C\\ 
D & 2 $\cdot$ $10^{15} \,\,\nicefrac{\text{ions}}{\text{cm}^2}$ & $5.5 \times 10^{-6}$ Torr& $5.5 \times 10^{-8}$ Torr&$3.1 \times 10^{-7}$ Torr& $525^{\circ}$C\\ 
E & 4 $\cdot$ $10^{15} \,\,\nicefrac{\text{ions}}{\text{cm}^2}$ & $5.5 \times 10^{-6}$ Torr& $1.8 \times 10^{-7}$ Torr&---& $370^{\circ}$C\\ 
Series F & 4 $\cdot$ $10^{15} \,\,\nicefrac{\text{ions}}{\text{cm}^2}$ & $5.5 \times 10^{-6}$ Torr& $1.8 \times 10^{-7}$ Torr&---& $330^{\circ}$C to $410^{\circ}$C\\ 
Series G & 4 $\cdot$ $10^{15} \,\,\nicefrac{\text{ions}}{\text{cm}^2}$ & $4.0 \times 10^{-6}$ Torr& $1.3 \times 10^{-7}$ Torr&---&$380^{\circ}$C\\ 
H & 2 $\cdot$ $10^{15} \,\,\nicefrac{\text{ions}}{\text{cm}^2}$ & $5.5 \times 10^{-6}$ Torr& $1.3 \times 10^{-7}$ Torr&---&$375^{\circ}$C\\ 
\bottomrule
\end{tabular}
\label{tab:proben}
\end{table*}
The advantages of this technique are that there is no need to use any mask, resist or similar preparation details to create single NW growth in designed pattern or even density variations 
and in addition the focused ion beam enables the use of different catalyst materials, depending on the Liquid Metal Ion Source (LMIS) \cite{Bischoff.2005}.\\
\\
In the following we will first report properties of non site controlled FIB induced NWs. The dependence on temperature, arsenic species and implantation fluence will be in investigated in terms of morphology. Additionally, the crystallization and growth direction of self assembled NWs will be presented. Then, we will focus on the properties of site controlled NWs. Especially the numbers of ions per spot to nucleate a single NW and the growth direction will be examined.

\kapitell{II. Growth}
For ion implantation and the NW growth a single-side polished (111) B arsenic terminated GaAs wafer is used as a substrate. The wafer is epiready to avoid further preprocessing. \\
To implant the Au$^{+}$ ions a LMIS, filled with a AuGa- or AuSiBe-alloy is used. 
The LMIS are produced by using eutectic alloys and a wire made of a nonreactive element, high melting point, e.g. tungsten \cite{Mazarov.2008,Orloff.2003}.\\
To implant the $\text{Au}^+$ ions, an acceleration voltage of $\text{U}_{\text{acc}}=30 \text{ keV}$ is used. The beam diameter is around 100 nm and 
 the resulting fluence is as descripted for the different samples.
\\
Before introducing the sample into the MBE chamber, it is degassed for 45 minutes at 150$^{\circ}$C in the load-lock chamber to reduce contaminations.\\
To create the catalyst droplet, the substrate temperature is increased to \tsub{550} for 5 minute in the growth chamber at UHV conditions. The implanted Au forms alloy droplets with the substrate material \cite{Mohammad.2008, Hannon.2006}, which are used for the growth of the NWs via VSS mechanism.\\ 
To initiate the InAs NW growth, the substrate temperature is reduced to growth temperature ranging from \tsub{330} to \tsub{430} and the As valve is opened with a beam equivalent pressure (BEP) from \BEP{As}{4.0}{-6} to \BEP{As}{6.0}{-6}. The In/Ga shutters are opened with a BEP from \BEP{In}{5.5}{-8} to \BEP{In}{2.0}{-7} and a \BEP{Ga}{3.1}{-7}, depending on the grown nanowire material. The As pressure and species are controlled by a valved cracker cell. The average growth time is one hour, terminated by closing the group III shutters and reducing $T_{Subs.}$, which results in 5 $\upmu$m to 20 $\upmu$m long NWs, depending on the fluxes and growth temperatures.\\
An overview about the implantation and growth parameters for the described samples is found in Tab. \ref{tab:proben}. 
\\
The NWs are investigated as-grown by a scanning electron microscope (SEM) for all samples. Microstructural characterization was performed using transmission electron microscopy (TEM) on a Tecnai F20 G2 operating at 200 kV. For TEM sample preparation, the NWs were deposited on a TEM-grid.
In order to transfer the InAs NWs, we put a small amount of isopropanol on the grown NWs. By soft scratching on the surface, the NWs are released from the substrate and suspended into the isopropanol which then is charged into a syringe and extruded on the TEM-grid and dried for 15 minutes. \\

\begin{figure}[tb]
\includegraphics[width=0.49 \textwidth]{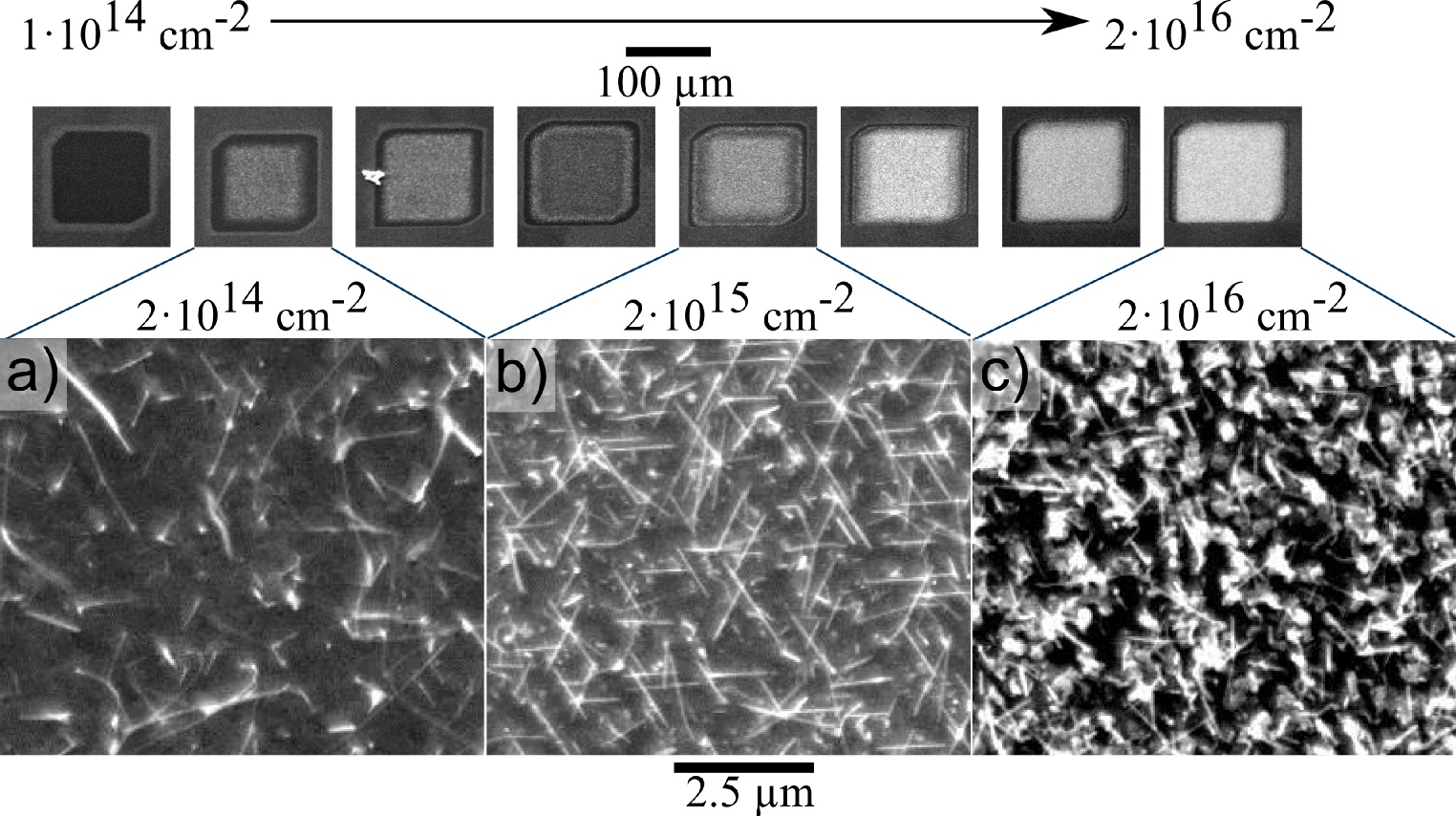}
\captionof{figure}{SEM pictures of NW fields grown on the sample A with different implanting fluences. With increasing fluence, the density of the NW increases until a critical density for cluster-forming is reached.}
\label{fig:areafluence}
\end{figure}

\begin{figure}[tb]
\includegraphics[width=0.49 \textwidth]{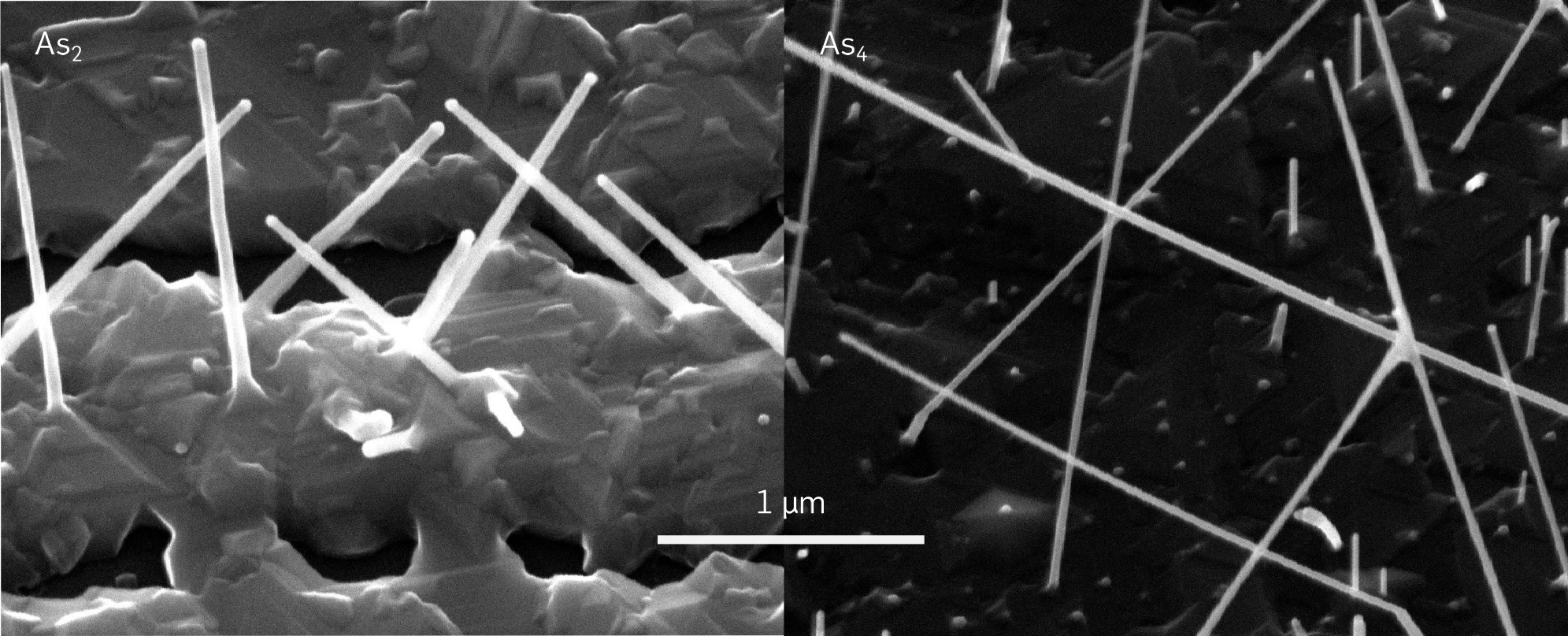}
\captionof{figure}{The image on the left shows a InAs nanowire sample of the series G with a higher As$_2$ to As$_4$ ratio, while on the right the As$_2$ to As$_4$ ratio was lowered. With a lower As$_2$ to As$_4$ ratio the length of the nanowires is increased, while the diameter is reduced. }
\label{fig:arsenic}
\end{figure}

\begin{figure}[tb]
\includegraphics[width=0.49 \textwidth]{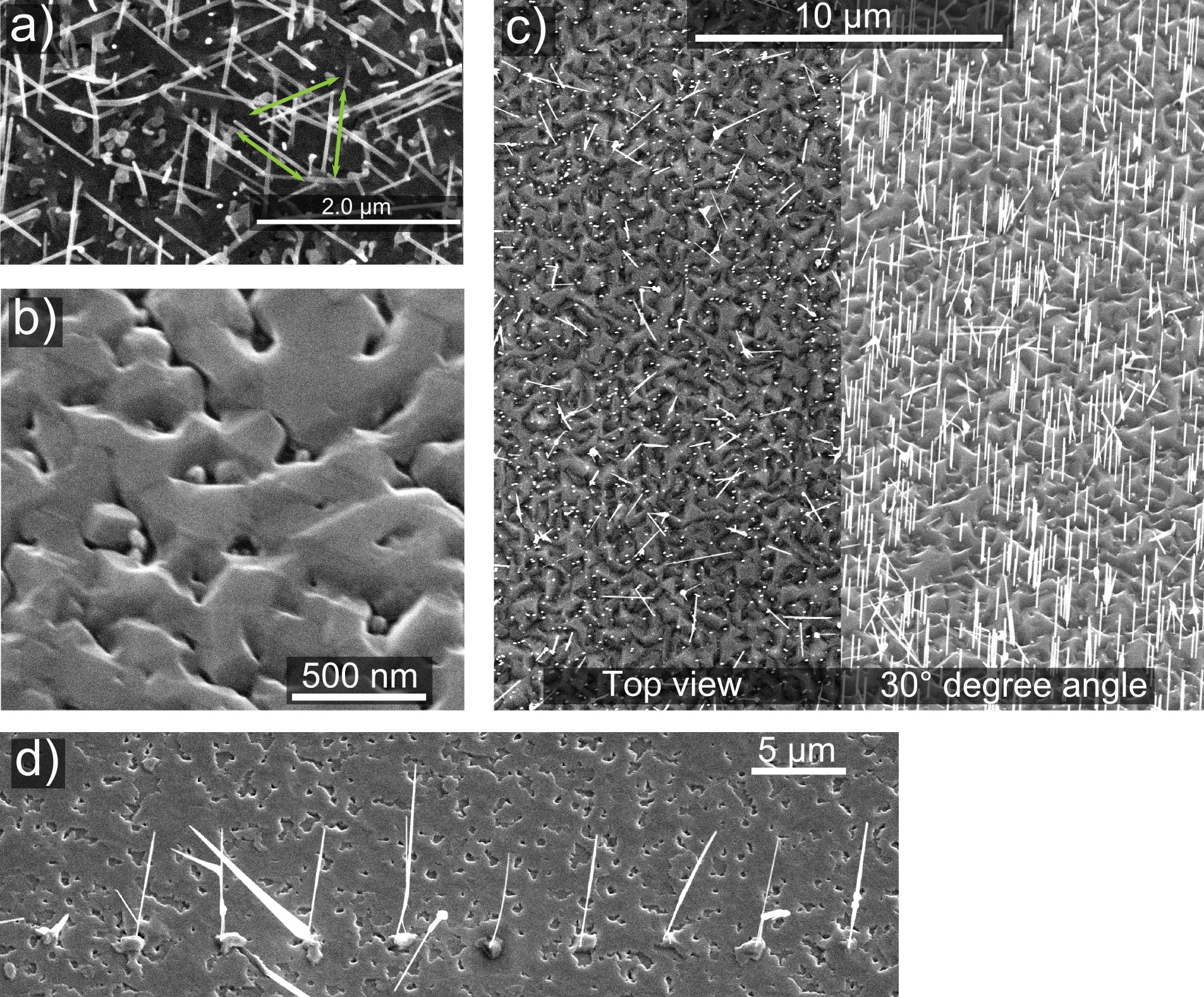}
\captionof{figure}{a) Top view of the InAs NW sample C. The projection of the NW growth direction got a 60$^{\circ}$ angle. b) close up of a pyramidal structure grown by InAs on the GaAs(111)B surface on sample E. c) In$_{0.15}$Ga$_{0.85}$As NW sample D, pictures taken from top view and under a 30$^{\circ}$ angle. With a high Ga amount most of the NWs grow in growth direction perpendicular to the surface. d) Single InAs NWs which grow more likely perpendicularly to the surface than area implanted InAs NWs.}
\label{fig:direction}
\end{figure}

\kapitell{III. Results}
To study the influence of the ion implantation, we have to distinguish between large-area and single-spot implantation.
In Fig. \ref{fig:areafluence}, the qualitative dependence of InAs NW density on the fluence of sample series A for area implantation is shown. A threshold fluence of $10^{14}\, \nicefrac{\text{ions}}{\text{cm}^2}$ could be identified at \tsub{370}, a BEP of \BEP{In}{1.8}{-7} and \BEP{As}{5.5}{-6}. With increasing fluence the density of the NWs increases until a critical point is reached. After reaching the maximal density around $1 \times 10^{16}\,{\text{cm}^{-2}}$ the NWs begin to form clusters on the surface like shown in Fig. \ref{fig:areafluence}c) \cite{Martelli.2011}. This results in a decreasing density of NWs due to the increasing density of the nanoclusters. \\
\begin{figure}[tb]
\includegraphics[width=0.49 \textwidth]{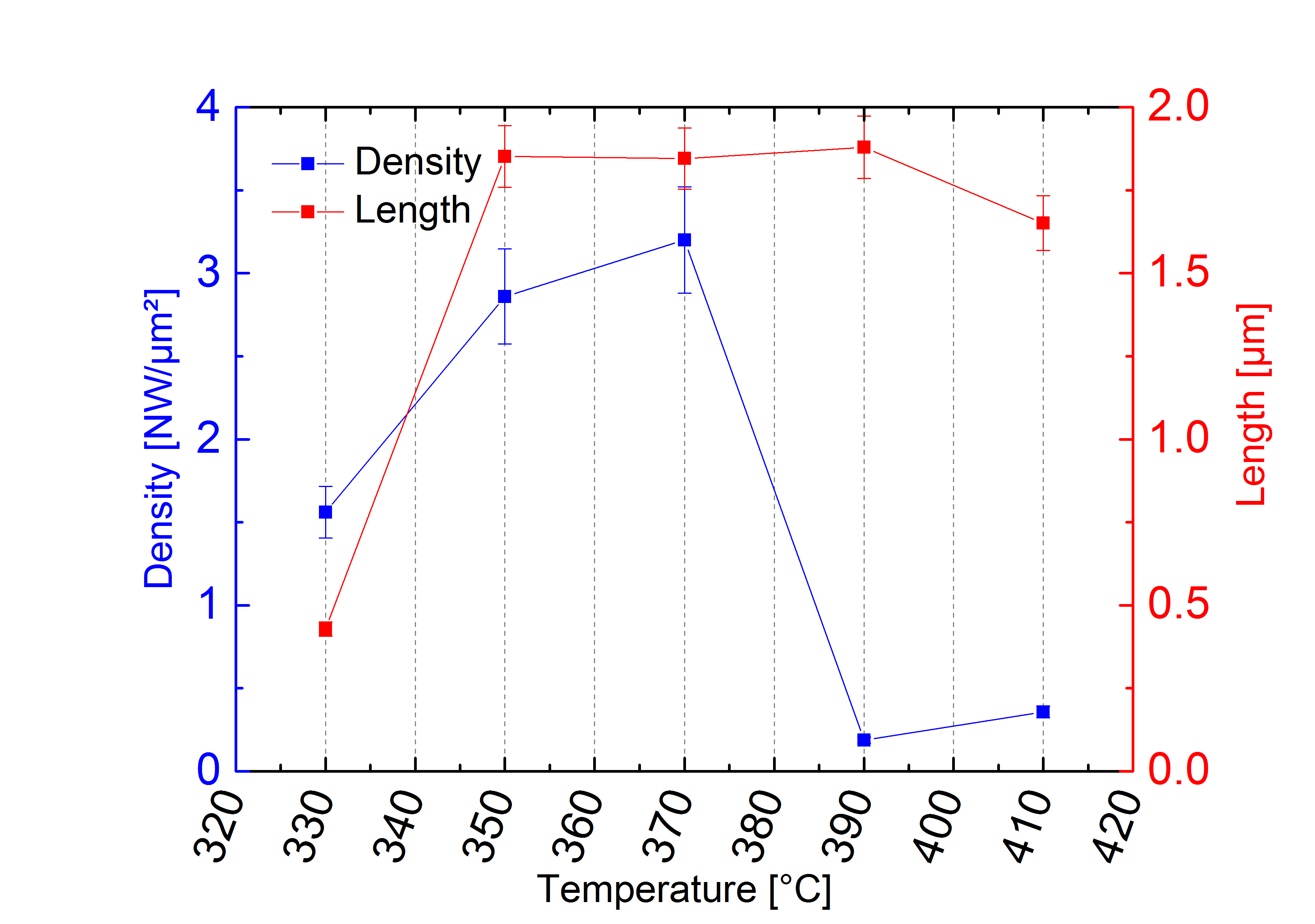}
\captionof{figure}{The graph shows as a function of the temperature the density and the length of the InAs nanowires for the sample series F. While growth was possible in a temperature range between 330$^{\circ}$C and 410$^{\circ}$C, an optimal growth temperature of 370$^{\circ}$C could be verified. }
\label{fig:temp}
\end{figure}\\
The morphology is given by the growth conditions and was examined in different approaches. 
Beside the In/As ratio - in the range of 3 to 5 - the substrate temperature $T_{\text{S}}$ has a major influence on the aspect ratio of NWs. Different samples were grown where the substrate temperature was varied between 330$^\circ$C and 410$^\circ$C.  
In Fig. \ref{fig:temp}, the density and length of the InAs NWs is shown for different $T_{\text{S}}$. At low temperatures in the range of 330$^\circ$C, the NWs start to grow. With increasing $T_{\text{S}}$, both the length and density of the NWs raise significantly. At $T_ {\text{S}} = 370^{\circ}$C, both the length and the density of the NWs have a maximum value. At higher temperatures, the density of the InAs NWs drops substantially and for substrate temperatures above 430\C, no InAs NW growth could be achieved. \\ 
\\
Another way to influence the morphology of the NWs is given by the used As species. The As ratio and pressure could be changed by using a valved cracker cell. For a higher As$_2$ to As$_4$ ratio, a cracker temperature of 700\C was used, while for a lower As$_2$ to $As_4$, ratio a cracker temperature of 500\C was sufficient.\\
 In Fig. \ref{fig:arsenic}, SEM images of two different samples from series G are shown. For the first sample, a higher As$_2$ to As$_4$ ratio was used (left image Fig.\ref{fig:arsenic}) which leads to a lower NW aspect ratio.  \\
The different length of the NWs could be explained by the higher sticking coefficient of As$_2$ \cite{Neave.1983}, which leads to a lower diffusion length $\lambda$ (see Fig.\ref{fig:vls}) of the group III adatoms and reduces the length of the InAs NWs corresponding to the VSS mechanism \cite{Madsen.2013,Mohammad.2008, Harmand.2005}. \\
By using a higher As$_4$ to As$_2$ ratio, the aspect ratio of the InAs NWs were increased to 300:1 which corresponds to a length up to 10 $\upmu$m with a diameter of 30 nm and below. For the higher As$_2$ to As$_4$ ratio, the length was reduced to 2 $\upmu$m while the diameter increased to 60 nm.
The growth directions of the NWs is not perpendicular to the surface for most of the NWs like shown in Fig.\ref{fig:direction} a). A high percentage of the InAs NWs grow in a projected 60$^\circ$ angle when looking onto the substrate.
\\
The reason for this might be the high lattice mismatch for layer growth of InAs ($a_{\text{InAs}}=6,058 \mathring{A}$ for a zincblende structure in F-43m group) and GaAs ($a_{\text{GaAs}}=5,653 \mathring{A}$ for a zincblende structure in F-43m group) \cite{Mikhailova.1996, Levinshtein.1996}. Therefore the planar growth mechanism is not as smooth as for GaAs epitaxial growth on the GaAs (111)B \cite{Hooper.1993}. 
The surface is roughend by triangular based pyramids right after the opening of the shutter, like shown in Fig. \ref{fig:direction} b). After forming the catalyst, the growth of the NWs does not start at the same time as the planar growth, but after a few minutes delay. 
Therefore, the [001] growth direction could be explained by the delayed growth of the nanowires on the side facets of the pyramids on the substrate. \\
Another hint is given by the proportionality between the In/Ga ratio in InGaAs NWs and the amount of NWs in the [001] growth direction. In Fig. \ref{fig:direction} c), there is a SEM image of In$_{0.17}$Ga$_{0.83}$As NWs taken from the top and with an angle of 30$^\circ$ with respect to the surface normal to estimate the growth direction of the NWs. Due to the high Ga amount a high percentage of the NWs grow orthogonally to the surface, in comparison to the pure InAs NWs, where only a few NWs grow orthogonal to the surface in area implantation. For pure GaAs NWs nearly all of the NWs grew orthogonal to the substrate. To conclude: A higher In ratio results in fewer perpendicularly grown NWs, which could be explained by a lower density of pyramides on the substrate when the NW growth starts.
\\
When using single spot implantation for single NW growth, the percentage of InAs NWs growing orthogonally is increasing significantly like shown in Fig. \ref{fig:direction} d).
\begin{figure}[tb]
\includegraphics[width=0.49 \textwidth]{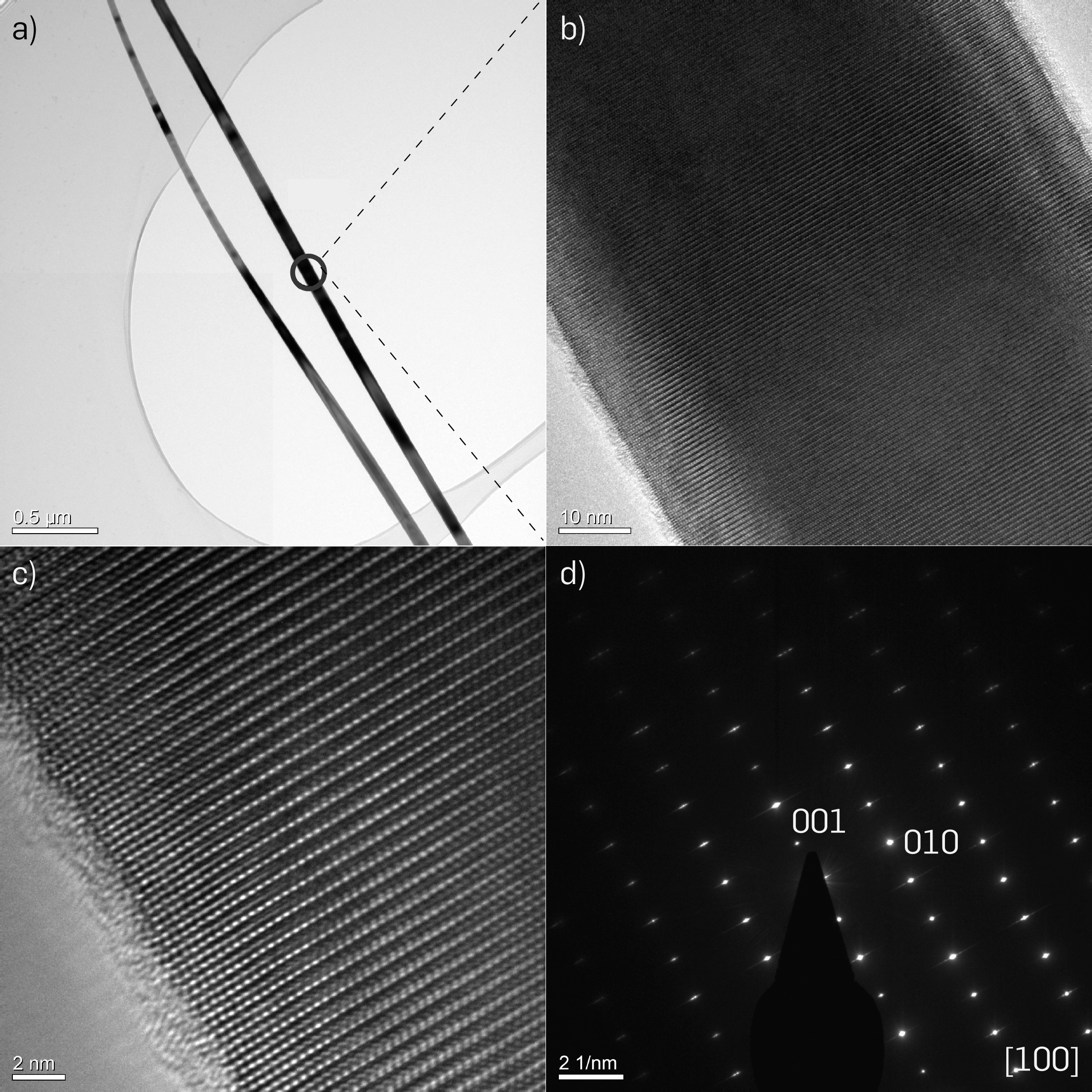}
\captionof{figure}{a) - c) TEM images of the NW sample H for different magnifications. d) TEM selected area electron diffraction pattern obtained from the area marked with a grey circle in a). The diffraction pattern reveals the hexagonal wurtzite structure of the NW. The microstructure of the InAs nanowire is mostly defect free.}
\label{fig:tem}
\end{figure}
\begin{figure}[tb]
\includegraphics[width=0.49 \textwidth]{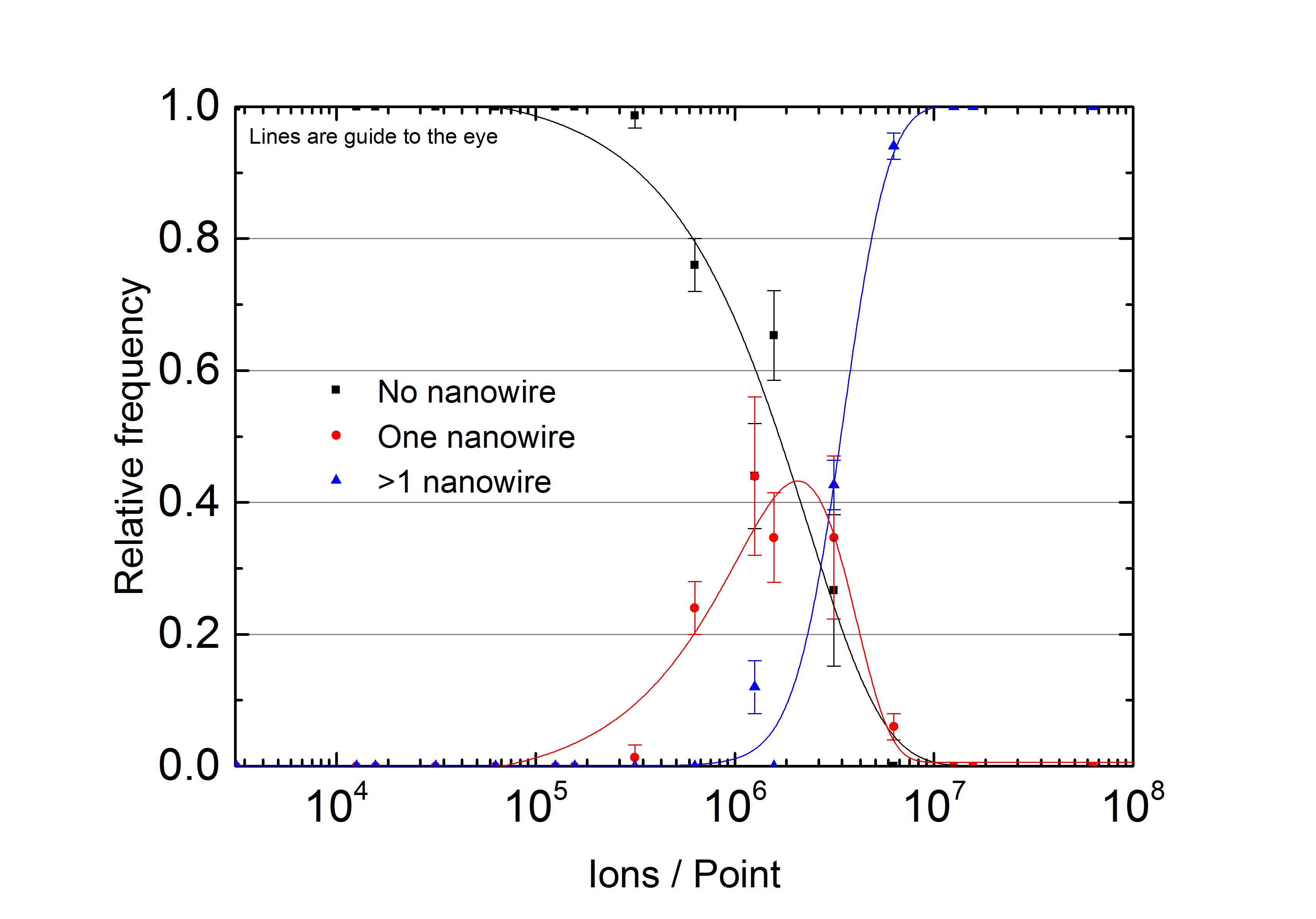}
\captionof{figure}{The graph shows the relative appearance of zero, one and more than one NW per implantation point depending on the implanting fluence grown on sample B. For single NW growth a optimal ion quantity of $2 \times 10^6 \, \nicefrac{\text{ions}}{\text{point}} $ for the given setup was determined.  }
\label{fig:singleNW}
\end{figure}\\
The microstructure of the InAs NWs was investigated by TEM. The studied NWs of sample H in Fig. \ref{fig:tem} were grown at \BEP{As}{5.5}{-6}, \BEP{In}{2.0}{-7} and \tsub{370} with a growth time of one hour and an implanting fluence of $2 \times 10^{15}\, \nicefrac{\text{ions}}{\text{cm}^2}$, like described in Tab. \ref{tab:proben}. The NWs got a diameter around 50 nm and a length between 4 $\upmu$m and 6 $\upmu$m.\\
A typical TEM bright field image and high resolution TEM images of an InAs NW are shown in Fig.\ref{fig:tem} a) - c), respectively. The NW appears with a homogeneous contrast and within the NW, no stacking faults or grain boundaries are observed on several NWs while scanning from the bottom to an area close under the droplet. Right under the droplet, a few crystal defects could be observed. Fig.\ref{fig:tem}d) shows a TEM selected area electron diffraction pattern obtained from the area marked with a white circle in Fig.\ref{fig:tem}a). Analysis of the diffraction pattern reveals that the crystal structure of the NW is of the hexagonal wurtzite type and not of zincblende structure, known from bulk InAs \cite{Mikhailova.1996}.   
\\
For the single NW growth, the fluence is given by the implanted ions per implantation point. To gain single NWs, there is an optimal implanting fluence like shown in Fig. \ref{fig:singleNW}. \\
For a low implanting fluence below $10^5 \, \nicefrac{\text{ions}}{\text{point}}$, no NW growth could be achieved. With increasing fluence, the chance of creating a single NW from one implantation point is increasing till a maximum effectivity of 45 \% is reached at $2 \times 10^6 \, \nicefrac{\text{ions}}{\text{point}} $. With even higher fluences, the chance of getting two or more NWs out of a single implantation point is increasing. 
\\
\kapitell{IV. Conclusion}
Induction of InAs NW growth due to catalyst implantation by focused ion beam was established. With this method, a simple and clean single step preparation for self assembled and site-controlled NW growth was achieved. Single InAs NW growth was shown with an effectivity of $\approx 45\%$ depending on the implantation fluence. For increasing fluence, the number of NWs induced from one implantation point is rising. For area implantation, a threshold fluence of  $10^{14} \nicefrac{\text{ions}}{\text{cm}^2}$ was found for the used FIB system.\\
The influence of different growth parameters on the InAs NW growth was reported. The optimal growth temperature was found to be $370^{\circ}$C where both, the aspect ratio and the density of the InAs NWs have a maximum value. For higher aspect ratios, a lower As$_2$ to As$_4$ ratio has to be used, which results in thinner and longer NWs. \\
With optimized the growth parameters and with focused ion implantation it was possible to grow pure monocrystalline InAs NWs which show no stacking faults and aspect ratios up to 300:1.
\kapitell{Acknowledgment}
Authors would like to thank for support through the Materials Research Department of Ruhr-Universität Bochum. Addionally we acknowledge gratefully support of Mercur  Pr-2013-0001, DFG-TRR160,  BMBF - Q.com-H  16KIS0109, and the DFH/UFA  CDFA-05-06.
\bibliography{paper}{}
\end{document}